\documentstyle[aps]{revtex}

\begin{document}
\title{''Pion laser'' phenomenon and other possible signatures of the DCC at
RHIC
and LHC energies}
\author{S.V. Akkelin and Yu.M. Sinyukov\thanks{%
e-mail: sinyukov@bitp.kiev.ua}}
\address{Bogolyubov Institute for Theoretical Physics, \\
National Academy of Sciences of Ukraine, Kiev 143, UKRAINE}
\maketitle

\begin{abstract}
The decay of fireballs containing the disoriented chiral condensate (DCC) in
A+A collisions has been analyzed. We found that the high phase-space density
and probably the large final fireball volume are the favorable factors to
extract a DCC signal from the thermal background. Both of these factors are
expected to take place at RHIC and LHC energies. A coherent pion component
then can be observed in pion spectra, in Bose-Einstein correlations and in
distribution of the ratio of neutral to total pions.
\end{abstract}



\section{Introduction}

The main goal of the present and future heavy ion experiments is to discover
the new states of the matter associated with chiral and deconfinement phase
transitions. The former may be accompanied by the presence of the
disoriented chiral condensate (DCC), i.e. correlated region of space-time
wherein the quark condensate is chirally rotated from its usual orientation
in isotopic space (see, for example, the reviews \cite{Bjorken}). The
characteristic momentum spread in the DCC is fixed by the inverse size of
the region. Recent experimental searches of the DCC signals exploit the
overall distribution of the ratio of neutral to total pions without momentum
cut as an experimental matter it is not a simple to select low momenta
neutral pions. Up to now a clear evidence of the DCC formation has not
found. Perhaps, the reason is that the DCC signal is strongly contaminated
by the thermal background. Therefore the combine strategy using the analysis
of the one- particle inclusive spectra and correlations function of charged
pions on soft momenta region could be effective tool to distinguish DCC..

The ability to identify DCC against the background critically depends on the
size and phase-space densities of the fireballs. For high enough phase-space
densities, which can be reached at RHIC and LHC, the ''pion laser''
phenomenon could take place. The conception of ''pion laser'' was introduced
in one of the first papers \cite{Pratt} devoted to the problem of
symmetrization of pion emission amplitudes at large phase-space densities in
finite source. However, the model of independent factorized sources (MIFS) 
\cite{Pratt,Lednicky,Csorgo,Zajc} leads to the completely chaotic radiation
without coherence. The following arguments can explain such a property of
the model. The intensive radiation in the MIFS arising at critical value of
parameters corresponds to the high-temperature Bose-Einstein (BE)
condensation of ideal gas in a finite system \cite{Sinyukov}. For infinite
homogeneous systems, the BE condensate in ideal gas is associated with the
phase transition where the coherent condensate wave function can be
considered as an order parameter. This phase transition is conditioned by
spontaneous breaking a gauge symmetry. To describe it, one should input a
fictitious source into the Hamiltonian, go to the thermodynamic limit and 
{\it after that} switch off the source \cite{Bogoliubov1}. This
non-commutative limit procedure leads to the ground state with nonzero
expectation value of field, i.e., to the coherent state. However, for
spatially inhomogeneous (effectively finite) systems with the finite number
of particles, there is no mechanism which results in spontaneous symmetry
breaking down. In fact, for finite systems the Hamiltonian of the ''pion
laser'' model has to include an evident symmetry breaking term, for
instance, an interaction with a classical source. Then the phenomenon of
''pion laser'' might be regarded as an effect of weak coherent signal
strengthening in dense boson thermal environment. The decay of the DCC is
one of the possible mechanisms for classical source to be introduced into
the Hamiltonian.

This paper is aimed at finding some experimentally observed consequences of
the hypothesis that the DCC can be associated with classical sources
interacting with quanta of dense bosonic medium. It is worthy to note that
though the whole picture developed is a simplified version of more realistic
scenario, it is believed to include some essential features of soft pion
production in future experiments at the next generation of heavy ion
colliders: RHIC and LHC.

In Section~II the space-time evolution of the DCC field is described by the
equations of motion of the linear sigma model. Since pions and sigma
particles are in a heat bath, the self-consistent Hartree, or mean-field,
approximation is appropriate for the model description (see, e.g., \cite
{Randrup}). According to Ref. \cite{Koch}, the interaction with the
surrounding heat bath is supposed to result in the finite lifetime of
classical pion field, and the corresponding decay width is introduced. It
leads to an additional term, describing the interaction of quasipions with
classical source, in the quasiparticle Hamiltonian. Actually, the DCC exists
in a finite hot inhomogeneous system that is surrounded by the normal vacuum
and hydrodynamically expands with a finite velocity. Thus, it should be
stressed that the problem of finding solutions of the self-consistent
equations is a very complicated one. Here, we use only some general
properties of such solutions.

In Section~III the quasiequilibrium stage of the evolution is described by
the thermal density matrix for inhomogeneous, effectively finite fireball.
The DCC decay in thermal bath can violate the chemical equilibrium for
quasipions, and for this reason the corresponding chemical potentials enters
the density matrix.

When the system loses the local thermal equilibrium, the description based
on the thermal density matrix is no longer adequate. In Section~IV we
describe the dynamic system evolution at this freeze-out stage by similar
form of the Hamiltonian as one at the equilibrium stage (the method is close
to \cite{Boyanovsky}). The main parameters of the evolution such as the
decay width and the effective time dependent pion masses cannot be, in
principle, derived from the self-consistent equations at this stage, they
are phenomenologic parameters and differ from themselves at the
quasiequilibrium stage. Although we do not consider here other higher order
interactions in the system which could result in the BE condensate
superfluidity (see, e.g., \cite{Bogoliubov2}), we suppose, however, that the
freeze-out at high densities can be explained on the basis of the
superfluidity of the condensate component of pion liquid.

Results are presented in Section~V where signatures of the DCC decay for
pion spectra, for HBT correlations and $N_{\pi _0}/N_{\pi _{tot}}$
distribution are analyzed. In Section~VI we summarize our results and
conclusions.

\section{Model of the DCC decay}

The linear sigma model describes the O(4) chiral field $\varphi _a=(\sigma ,%
\overrightarrow{\pi }{\bf )}$ by means of the Lagrangian with a simple
nonlinear self-interaction 
\begin{equation}
{\cal L}=\frac 12\partial _\mu \varphi ^a\partial ^\mu \varphi ^a-\frac 
\lambda 4(\varphi ^a\varphi ^a-v^2)^2+h\sigma .  \label{Lsigma}
\end{equation}
For the simplicity, let us suppose that in the Hartree approximation there
are no cross correlations between different components of the field $\varphi
_a.$ Then the equations of motion for pion $"i"$ component of the order
parameter (classical field) $\varphi _{c,i}=\left\langle \varphi
_i\right\rangle $ and for the field fluctuations (quantum quasiparticles) $%
\varphi _{q,i}$ can be obtained from the Hartree equation for the pion field 
$\varphi _i=\varphi _{q,i}+\varphi _{c,i}$ (see, e.g., \cite{Chang}) 
\begin{equation}
\partial _\mu \partial ^\mu \varphi _i(x)+m_i^2(x)\varphi _i(x)=2\lambda
\left\langle \varphi _i(x)\right\rangle ^3,  \label{em}
\end{equation}
here $m_i^2(x)$ depends on the field averages. Indeed, Eq. (\ref{em}) can be
split into two equations 
\begin{equation}
\partial _\mu \partial ^\mu \varphi _{c,i}+m_{c,i}^2\varphi _{c,i}=-f_{c,i},%
\text{ \ \ }\partial _\mu \partial ^\mu \varphi _{q,i}+m_i^2\varphi
_{q,i}=f_{c,i},\text{ \ \ }m_{c,i}^2\equiv m_i^2-2\lambda \varphi _{c,i}^2,
\label{eq1}
\end{equation}
where $f_{c,i}(x)=0$ corresponds to the mean-field prescription. Eq. (\ref
{eq1}) for quasiparticles can be derived from the Hamiltonian $%
H_{q,i}(t)=\int d^3r{\cal H}_{q,i}(x),$ 
\begin{equation}
{\cal H}_{q,i}(x)=\frac 12(\pi _{q,i}^2(x)+({\bf \nabla }\varphi
_{q,i}(x))^2+m_i^2(x)\varphi _{q,i}^2(x))-f_{c,i}(x)\varphi _{q,i}(x),
\label{Hq-res}
\end{equation}
where $\pi _{q,i}(x)=\stackrel{\cdot }{\varphi }_{q,i}(x)$ is the
canonically conjugated momentum.

We suppose that in the mean-field approximation the classical field $\varphi
_{c,i}=$ $\widetilde{\varphi }_{c,i}({\bf r},t)=n_i\widetilde{\varphi }_c(%
{\bf r},t)$ is localized in a hot and dense region and is a quasistationary
state (its energy $\omega _c(t)$ changes adiabatically): 
\begin{equation}
\text{ }\widetilde{\varphi }_c({\bf r},t)=\widetilde{\psi }_c^{*}({\bf r,t})+%
\widetilde{\psi }_c({\bf r,t}),\text{ }\widetilde{\psi }_c({\bf r,t})=%
\widetilde{\psi }_c({\bf r})\exp (-i\int\limits_{t_0}^tdt^{\prime }\omega
_c(t^{\prime })),  \label{eq3}
\end{equation}
where $n_i(\theta ,\phi )$ is the $"i"$ component of the randomly oriented
3-vector ${\bf n}$, ${\bf n}^2=1,$ we suppose that ${\bf n}$ does not depend
on $x.$ The orientation of the vector ${\bf n}$ can be chosen arbitrarily
due to the symmetry of the sigma model Lagrangian.

The interaction of the classical field with thermal quasiparticles leads to
the finite lifetime of the DCC \cite{Koch}. After the decay starts, the
quantum level is shifted by $\delta \omega $ and the decay width $\Gamma $
appears in the perturbation theory approximation. Let us neglect $\delta
\omega $ and suppose that $\Gamma (t)=\Gamma _1\Theta (t_f-t)+\Gamma
_2\Theta (t-t_f)$, $\Gamma _1\ll m_i(x)$ , $\Gamma _2=\Gamma _1$ or $\Gamma
_2\gg \Gamma _1.$ As it is discussed later, the last prescription
corresponds to a rapid disintegration of the system at $t>t_{f\text{ }}$
just below the ''thermal freeze-out' temperature $T_f=T(t_f)$. Then taking
the decay of the classical field into account, we have: 
\begin{equation}
\varphi _{c,i}(x)=\exp (-\frac 12\int\limits_{t_0}^tdt^{\prime }\Gamma
(t^{\prime }))\widetilde{\varphi }_{c,i}({\bf r},t).  \label{cf}
\end{equation}
By using the first equation in (\ref{eq1}) and (\ref{cf}), one can find the
deviation from pure mean field prescription in the case of DCC decay: 
\begin{equation}
\text{ }f_{c,i}(x)=n_if_c(x)=-\stackrel{..}{\varphi }_{c,i}(x)+\exp (-\frac 1%
2\int\limits_{t_0}^tdt^{\prime }\Gamma (t^{\prime }))\stackrel{\cdot \text{ }%
\cdot }{\widetilde{\varphi }}_{c,i}(x)\neq 0.  \label{cf1}
\end{equation}

\section{Thermal quasiequilibrium stage}

At the initial stage $(t_0\leq t\leq t_f)$, the system is in a thermal
equilibrium and is described by the quasipion Hamiltonians $H_{q,i}(t)$ (\ref
{Hq-res}) and the statistical operator $\rho _{{\bf n}},$ 
\begin{equation}
\rho _{{\bf n}}Sp\rho _{{\bf n}}=e^{-\sum_iH_{q,i}^{^{\prime }}(t)/T},\text{ 
}H_{q,i}^{^{\prime }}(t)=\int d^3r[{\cal H}_{q,i}(x)-\mu _i{\cal J}%
_{q,i}(x)-\mu ^{\prime }(x){\cal J}_{q,i}^{\prime }(x)].  \label{eq7}
\end{equation}
Here $\mu _i$ are the chemical potentials which are responsible for
violation of the chemical equilibrium and ${\cal J}_{q,i}(x)=\varphi
_{q,i}^{(+)}(x)\stackrel{\longleftrightarrow }{\frac \partial {\partial t}}%
\varphi _{q,i}^{(-)}(x)$ where $\varphi _{q,i}^{(+)}$ and $\varphi
_{q,i}^{(-)}$ are the positive and negative field components, respectively.
The chemical potentials appear due to the decay of classical field which
accompanies the creation of new quasiparticles with their subsequent fast
thermalization in the dense environment. To guarantee a finiteness of the
system, we include the effective phenomenologic ''chemical potential'' $\mu
^{\prime }(x)$ in the statistical operator: 
\begin{equation}
\mu ^{\prime }(x){\cal J}_{q,i}^{\prime }(x)=-\frac 12\varphi
_{q,i}^2(x)V(r),\text{ \ \ }V(r)=m^2(t_f)\omega ^2r^2\equiv \frac{m(t_f)T_f}{%
R^2}r^2.  \label{eq22}
\end{equation}
Let us suppose that all $\mu _i$ are equal to $\mu $ and $%
m_i(x)=m_i(t)\simeq m(t)$. We can disregard the coordinate dependence of
masses in Eq. (\ref{eq7}) because the ''trapping potential'' $V(r)$ acts as
a strong cut-off factor. Note, however, that $V(r)$ is not real potential:
it is not a part of the Hamiltonian $H_{q,i}(t)$. At $\mu =0$ the quadratic
part of $H_{q,i}^{^{\prime }}$ in Eq. (\ref{eq7}) can be diagonalized at $%
t=t_{f\text{ }}$ in the basis $\xi _n({\bf r})\equiv \xi _{n_1}(x)\xi
_{n_2}(y)\xi _{n_3}(z),$ 
\begin{equation}
\varepsilon _n^2(t_f)\xi _n({\bf r})=(-\Delta +V(r)+m^2(t_f))\xi _n({\bf r}),
\label{eq23}
\end{equation}
with the creation and annihilation operators $a_{n,i}^{\dagger }(t_f)$, $%
a_{n,i}.(t_f)$, where (see, e.g. \cite{Landau}) 
\begin{eqnarray}
\xi _{n_1}(x) &=&\frac{H_{n_1}(x\sqrt{m(t_{in})\omega })\exp
(-x^2m(t_f)\omega /2)}{((n_1!)2^{n_1}\sqrt{\pi /m(t_f)\omega })^{1/2}},
\label{eq23.1} \\
\text{ }\varepsilon _n^2(t_f) &=&2m(t_f)\omega (n_1+n_2+n_3+\frac 32%
)+m^2(t_f).  \label{eq23.2}
\end{eqnarray}
At the same time the quadratic part of the quasiparticle Hamiltonian $%
H_{q,i}(t)$ can be diagonalized in the momentum representation with the
operators $a_i({\bf k,}t_f)$ and $a_i^{\dagger }({\bf k,}t_f)$. The relation
between these representations is given by the Bogolyubov transformation with
the coefficients $A_n({\bf k})$ and $B_n({\bf k}):$ 
\begin{eqnarray}
\left( 
\begin{array}{c}
a_i({\bf k},t_f) \\ 
a_i^{\dag }({\bf k},t_f)
\end{array}
\right)  &=&\sum_n\left( 
\begin{array}{cc}
A_n({\bf k}) & B_n({\bf k}) \\ 
B_n({\bf k}) & A_n({\bf k})
\end{array}
\right) \left( 
\begin{array}{c}
a_{n,i}(t_f) \\ 
a_{n,i}^{\dagger }(t_f)
\end{array}
\right) ,  \label{eq24} \\
A_n({\bf k}) &=&\frac 12\left( \sqrt{\frac{\varepsilon _k(t_f)}{\varepsilon
_n(t_f)}}+\sqrt{\frac{\varepsilon _n(t_f)}{\varepsilon _k(t_f)}}\right) \xi
_n({\bf k}),  \label{eq24.1} \\
B_n({\bf k}) &=&\frac 12\left( \sqrt{\frac{\varepsilon _k(t_f)}{\varepsilon
_n(t_f)}}-\sqrt{\frac{\varepsilon _n(t_f)}{\varepsilon _k(t_f)}}\right) \xi
_n(-{\bf k}).  \label{eq24.2}
\end{eqnarray}
Eqs. (\ref{eq24}) - (\ref{eq24.2}) can be simply derived if the canonical
field variables (coordinate and momentum) are written in terms of the
creation and annihilation operators in different representations. The
coefficients $A_n({\bf k})$ and $B_n({\bf k})$ satisfy to the condition $%
\sum_n(A_n({\bf k}_1)A_n({\bf k}_2)-B_n({\bf k}_1)B_n({\bf k}_2))=\delta (%
{\bf k}_1-{\bf k}_2).$ The last relation means that it is a canonical
transformation. It is easy to see that the Fock spaces for $a_{n,i}^{\dagger
}(t_f)$ and $a_i^{\dagger }({\bf k},t_f)$ are the unitary nonequivalent
representations of the canonical relations. Physical vacuum is the ground
state of the Hamiltonian $H_{q,i}(t)$ and is defined as $a_i({\bf k}%
,t_f)\left| 0\right\rangle =0$, whereas unphysical vacuum of the
''Hamiltonian'' $H_{q,i}^{^{\prime }}$ is defined by $a_{n,i}(t_f)\left|
0^{\prime }\right\rangle =0$. So, one should renormalizes the ground state
of the statistical operator:

\begin{equation}
\left\langle ...\right\rangle _{ren}=\left\langle ...\right\rangle
-\left\langle 0^{\prime }\left| ...\right| 0^{\prime }\right\rangle ,\text{ }%
\left\langle ...\right\rangle =Sp(...\rho ).  \label{renorm}
\end{equation}
Below we omit the index $"ren"$. At large $R\gg \frac 1{m(t_f)}\sqrt{\frac{%
T_f}{m(t_f)}}$ one can calculate the thermal averages of the operators $a(%
{\bf k},t_f),$ $a^{\dag }({\bf k},t_f)$ in the approximation: $A_n({\bf k}%
)\rightarrow \xi _n({\bf k}),$ $B_n({\bf k})\rightarrow 0.$ Then the thermal
averages at given ${\bf n}$ in the general case $\mu \neq 0$ are 
\begin{equation}
\left\langle a_i^{\dagger }({\bf p,}t_f)a_i({\bf p},t_f)\right\rangle _{{\bf %
n}}=N_{th}({\bf p})+\left| \left\langle a_i({\bf p},t_f)\right\rangle _{{\bf %
n}}\right| ^2,  \label{eq31}
\end{equation}
where 
\begin{equation}
N_{th}({\bf p})=\sum_n\frac{\xi _n^2({\bf p})}{\exp (\varepsilon _n-\mu )-1},%
\text{ }\left\langle a_i({\bf p},t_f)\right\rangle _{{\bf n}%
}=n_i\sum_ng_n\xi _n({\bf p}),  \label{eq32}
\end{equation}
and\ $g_n=\int d^3r\xi _n({\bf r})f_c({\bf r},t_f)/((\varepsilon _n-\mu )%
\sqrt{2\varepsilon _n}).$

\section{Freeze-out stage in the Heisenberg representation}

At the final stage $(t_f<t\leq t_{out}<\infty )$, the system loses the
(local) thermal equilibrium, and the misaligned vacuum goes to the normal
vacuum with the lifetime $\Gamma _2\propto (t_{out}-t_f)^{-1}.$ Here $t_{out}
$ is the ''physical asymptotic time'' when all interactions are neglected
and $\varphi _{c,i}({\bf r},t_{out})\simeq 0,$ $m_i({\bf r},t_{out})\simeq
m_\pi $, $a_i({\bf p},t_{out})\equiv a_{i,out}({\bf p})\simeq a_{\pi _i}(%
{\bf p},t_{out})$, the system evolution is determined by the free pions
Hamiltonian. We suppose that the evolution of the system during the {\it %
continuous} freeze-out stage $t_f<t\leq t_{out}$ is governed by the
Hamiltonian (\ref{Hq-res}) with the effective masses $m_i^2(x),$ as is close
to \cite{Boyanovsky}. It should be noted that the coordinate dependence of
the effective mass in Eq. (\ref{Hq-res}) ensures a finiteness of
squeeze-states contributions to spectra. These contributions appear due to
the difference in quasiparticles masses at $t=t_f$ and $t=t_{out}$, they are
small either in the adiabatic approximation for the mass evolution or when $%
m_i({\bf r},t_f)\simeq m_i({\bf r},t_{out})$. For the sudden freeze-out
scenario we will use the second assumption: $m_i({\bf r},t_f)\simeq m_i({\bf %
r},t_{out})$. In these cases we can make a further simplification supposing $%
m_i(x)=m_i(t)\simeq m(t)$. Then using the Hamiltonian (\ref{Hq-res}), we get
the well-known Heisenberg equation's solution for quantum fields which
interact with classical source 
\begin{equation}
a_{i,out}({\bf p})\simeq a_i({\bf p,}t_f)+n_id_{{\bf p}}(t_{out}),
\label{eqH}
\end{equation}
where 
\begin{equation}
d_{{\bf p}}(t)=\frac i{\sqrt{2\varepsilon _{{\bf p}}(t_f)}}%
\int\limits_{t_f}^tdt^{\prime }{\bf \exp (}i\int\limits_{t_f}^{t^{\prime
}}dt^{\prime \prime }\varepsilon _{{\bf p}}(t^{\prime \prime }))f_c({\bf p,}%
t^{\prime }),\text{ }\varepsilon _p(t)=\sqrt{{\bf p}^2+m^2(t)}  \label{eq10}
\end{equation}
and $f_c({\bf p,}t)$ is the Fourier-transformed current $f_c(x)$ in (\ref
{cf1}).

In the case of the adiabatic freeze-out $\Gamma =\Gamma _1=\Gamma _2\ll m(t)$%
, we have $d_{{\bf p}}(t_{out})\simeq 0$ for all ${\bf p}$ if $\omega _c<m.$
If $\omega _c\geq m$ then the main contribution has the Breit-Wigner form
and reaches its maximal at some momentum ${\bf p}$ when $\varepsilon
_p(t)\simeq $ $\omega _c(t)$ at some ''average'' time point $t=\stackrel{\_}{%
t}$: 
\begin{equation}
d_{{\bf p}}(t_{out})\simeq -\frac \Gamma {\sqrt{2\varepsilon _{{\bf p}}(t_f)}%
}\frac{\omega _c(\stackrel{\_}{t})\widetilde{\psi }_c({\bf p,}t_f)}{%
i(\varepsilon _{{\bf p}}(\stackrel{\_}{t})-\omega _c(\stackrel{\_}{t}))-%
\frac \Gamma 2}\exp (-\frac \Gamma 2(t_f-t_0)),  \label{eq14}
\end{equation}
The sudden freeze-out case corresponds to $\Gamma _2\rightarrow \infty .$
Then we have 
\begin{equation}
d_{{\bf p}}(t_{out})\simeq \frac 1{\sqrt{2\varepsilon _p}}{\bf \{}%
\varepsilon _k\varphi _c({\bf p,}t_f)+i\stackrel{\cdot }{\varphi }_c(-{\bf p,%
}t_f)\},  \label{d-s}
\end{equation}
and in terms of pion field operators, Eq. (\ref{eqH}) can be rewritten as $%
a_{\pi _i}({\bf p,}t_{out})\simeq a_{\pi _i}({\bf p,}t_f).$

\section{Signatures of the DCC decay}

If we relate two stages through Eqs. (\ref{eq31}) and (\ref{eqH}), the
thermal averages for the operators $a_{i,out}({\bf p})$ at given ${\bf n}$
are written as: 
\begin{equation}
\left\langle a_{i,out}^{\dagger }({\bf p})a_{i,out}({\bf p})\right\rangle _{%
{\bf n}}=N_{th}({\bf p})+\left| \left\langle a_{i,out}({\bf p})\right\rangle
_{{\bf n}}\right| ^2,  \label{out-th-1}
\end{equation}
where 
\begin{equation}
\left\langle a_{i,out}({\bf p})\right\rangle _{{\bf n}}=n_i(\sum_ng_n\xi _n(%
{\bf p})+d_{{\bf p}}(t_{out})).  \label{out-th-2}
\end{equation}
To evaluate inclusive spectra, we should average over all the orientations $%
{\bf n}$. By taking into account that $n_{+,-}=\frac{\sin \theta }{\sqrt{2}}%
e^{\pm i\phi },$ \ $n_0=\cos \theta ,$ the one- particle inclusive spectra
for $\pi ^{+},\pi ^{-},\pi ^0$ pions are 
\begin{equation}
N_{\pi ^{+},\pi ^{-},\pi ^0}({\bf p})=\frac 1{4\pi }\int N_{+,-,0}({\bf p}%
)\sin \theta d\theta d\phi =N_{th}({\bf p})+\frac 13N_{coh}({\bf p}),
\label{one-in}
\end{equation}
where 
\begin{equation}
N_{+,-,0}({\bf p})=N_{th}({\bf p})+\left| n_{+,-,0}\right| ^2N_{coh}({\bf p}%
),\;N_{coh}({\bf p})=\left| \sum_ng_n\xi _n({\bf p})+d_{{\bf p}%
}(t_{out})\right| ^2.  \label{n-charge}
\end{equation}
In order to calculate two-particle inclusive spectra, one has to apply the
thermal Wick theorem for fixed ${\bf n}$ and then to average over ${\bf n}$.

To characterize the relative coherent contribution, we consider the ratio ''%
{\it signal to noise}'': 
\begin{equation}
D({\bf p})=\frac{\frac 13N_{coh}({\bf p})}{N_{th}({\bf p})}.  \label{rat-def}
\end{equation}
As it is already mentioned, the classical field is localized in a hot and
dense region. The Gaussian length scale of this region for rare gas is $%
\widetilde{R}^2\simeq R^2,$ while for the dense BE condensate it is $%
\widetilde{R}^2\simeq R_c^2=\frac R{2\sqrt{m(t_f)T_f}}$ -- which is easily
seen from $\xi _0({\bf r}).$ Let us suppose for the simplicity that the
classical field is localized within the corresponding scale $\widetilde{R}$
with a Gaussian spatial distribution: 
\begin{equation}
\widetilde{\psi }_c({\bf r,}t_f)=\frac \kappa {\widetilde{R}^{3/2}\sqrt{%
\omega _{c(t_f)}}}\exp (-r^2/4\widetilde{R}^2).  \label{psi-ap}
\end{equation}
If the classical field is large enough at the initial moment $t_{0\text{ }}$
and $\Gamma _1(t_f-t_0)>1$, the DCC decay at the first thermal stage can
lead to an overpopulation of quasiparticles and hence to appearance of the
chemical potential $\mu \neq 0$, as it was discussed before. If $\mu \simeq $
$\varepsilon _0$, then the average phase-space density is large enough, $%
n_{ps}=(N_{coh}+\frac 13N_{th})/\overline{p}\overline{r}$ $\gg 1$, and is
mainly determined by the coherent condensate contribution 
\begin{equation}
n_{ps,coh}\propto \kappa ^2\frac{\omega _c}{\varepsilon _0}\left( \frac{%
\Gamma _1}{(\varepsilon _0-\mu )}\right) ^2\exp (-\Gamma _1(t_f-t_0)).
\label{n-ph-coh}
\end{equation}
The ratio {\it ''signal to noise'' } can be obtained from (\ref{eq32}), (\ref
{n-charge}), (\ref{n-ph-coh}): 
\begin{eqnarray}
D({\bf p}) &\simeq &\frac{\frac 13g_0^2\xi _0^2({\bf p})}{\xi _0^2({\bf p}%
)/(\exp (\varepsilon _0-\mu )-1)}\propto   \nonumber  \label{eq36} \\
&\propto &\kappa \frac{\Gamma _1}{T_f}\sqrt{\frac{\omega _cn_{ps,coh}}{%
\varepsilon _0}}\exp (-\frac{\Gamma _1}2(t_f-t_0))\propto \kappa \sqrt{n_{ps}%
}.  \label{eq36}
\end{eqnarray}
The large phase-space density $n_{ps}\gg 1$ results in the large value $D(%
{\bf p})\gg 1$ due to the weak signal strengthening in an overpopulated
dense boson medium. This is a ''pion laser'' phenomenon. This effect is
supposed to be experimentally observed for pions in the soft energy region: $%
(p^0-m_\pi )<R^{-1}\ln n_{ps}$. In pion transverse energy spectra, this
effect reveals itself in an essential decrease of the spectrum slope at
small energies, $\frac 1{4m_\pi R_c^2}$ , compared to at high energies, $T_f.
$ The same effect observes also in a dense non-coherent condensate \cite
{Pratt,Lednicky,Csorgo}, \cite{Sinyukov} but the thermal freeze-out of such
a dense system can be hardly explained if there are no coherent attributes
(e.g., superfluidity).

If the overpopulation at the thermal stage does not arise ($\Gamma
_1(t_f-t_0)<1$) and therefore either $\mu \simeq 0$ or it is far from the
critical value $\mu =\varepsilon _0$, the condensate contribution is
negligibly small. In the non-relativistic approximation and at sufficiently
large volume, $R\sqrt{m(t_f)T}\gg 1$, the thermal part of the radiation can
be roughly described (at $R(\varepsilon _0-\mu )\sqrt{\frac T{m(t_f)}}\gg 1$
) as the BE gas \cite{Sinyukov}: 
\begin{equation}
N_{th}({\bf p})\simeq N_{gas}({\bf p})\equiv \frac 1{(2\pi )^3}\int \frac 1{%
\exp [({\bf p}^2/2m(t_f)+r^2T_f/2R^2+\varepsilon _0-\mu )/T_f]-1}d^3r.
\label{Ngas-ap}
\end{equation}
In this case the leading contribution to the coherent emission component $%
N_{coh}({\bf p})$ is $\left| d_{{\bf p}}(t_{out})\right| ^2$, and $d_{{\bf p}%
}(t_{out})$ reaches its maximum in the case of sudden freeze-out -- which we
use here as an illustration. From Eqs. (\ref{d-s}), (\ref{psi-ap}) and (\ref
{Ngas-ap}), we have 
\begin{equation}
D({\bf p})\simeq \frac{\frac 13\left| d_{{\bf p}}(t_{out})\right| ^2}{%
N_{gas}({\bf p})}\propto \kappa ^2\exp (-\Gamma _1(t_f-t_0))\exp (-{\bf p}%
^2(2R^2-\frac 1{2T_fm(t_f)})).  \label{D-rare}
\end{equation}
Some experimental effects can be observed for pions in the soft energy
region: $(p^0-m_\pi )<(4m_\pi R^2)^{-1}$ which is narrower than for ''pion
laser''. Within this region the inverse of energy spectrum slope, $(4m_\pi
R^2)^{-1}$, is considerably less than at high energies, $T_f.$ The
interferometry radius slightly decreases in this gas regime \cite{Sinyukov}.

The coherence is most directly connected with the intercept of correlation
function (CF). For the CF of $\pi ^{+}\pi ^{+}$-, $\pi ^{+}\pi ^{-}$- pairs
the intercepts are 
\begin{equation}
C^{++}({\bf p},{\bf 0})=2-\frac 45\left( \frac{D({\bf p})}{1+D({\bf p})}%
\right) ^2,\text{ }C^{+-}({\bf p},{\bf 0})=1+\frac 15\left( \frac{D({\bf p})%
}{1+D({\bf p})}\right) ^2.  \label{eq40}
\end{equation}
For $D({\bf p})\rightarrow \infty $ one get $C^{++}({\bf p},{\bf 0})=C^{+-}(%
{\bf p},{\bf 0})\rightarrow 1.2$, and the effective interferometry radius
squared, with 
\begin{equation}
C^{++}({\bf p},{\bf q}_{eff})-1=(C^{++}({\bf p},{\bf 0})-1)e^{--{\bf q}%
_{eff}^2R_{eff}^2},\text{ }{\bf q}_{eff}^2R_{eff}^2=1,  \label{eq37}
\end{equation}
drastically shrinks at condensation \cite{Sinyukov}: 
\begin{equation}
R_{eff}^2({\bf p})\simeq \frac{R_c^2}{1+2\ln (N_{\pi ^{+}}({\bf p})/N_{gas}(%
{\bf p}))}.  \label{eq43}
\end{equation}
Finally, let us discuss the problem of the DCC observation where the
distribution of the ratio of neutral to total pions $f=N_{\pi _0}/N_{\pi
_{tot}}$ is analyzed. It is easy to see from Eq. (\ref{n-charge}) that the
fixed ratio $f=N_{\pi _0}/N_{\pi _{tot}}$ corresponds to an averaging over $%
\phi $ at a fixed $\theta .$ Then using (\ref{n-charge}), we get 
\begin{equation}
f{\bf =}\frac{N_{\pi ^0,\theta }}{N_{\pi ^0,\theta }+N_{\pi ^{+},\theta
}+N_{\pi ^{-},\theta }}=\frac{\frac 13+D\cos ^2\theta }{1+D},  \label{eq38}
\end{equation}
where, for example, $N_{\pi ^0,\theta }=\int N_{\pi ^0,\theta }({\bf p}%
)d^3p, $ $N_{\pi ^0,\theta }({\bf p})=\frac 1{2\pi }\int N_0({\bf p})d\phi ,$
and $D=\frac 13\int N_{coh}({\bf p})d^3p/\int N_{th}({\bf p})d^3p\equiv 
\frac 13N_{coh}/N_{th}$. Therefore the $f$-distribution is 
\begin{equation}
P(f)=-\sin \theta \frac{d\theta }{df}=\frac 12\sqrt{\frac D{f(1+D)-\frac 13}}%
\frac{1+D}D  \label{eq42}
\end{equation}
in the interval $1/3(1+D)$ $<f<(1/3+D)/(1+D)$, and $P(f)=0$ beyond the above
interval. Note, that neither thermal nor coherent pion number fluctuations
does not taken into account in (\ref{eq42}). For a weak DCC\ signal, $D\ll 1$%
, the distribution is much more narrow than $P(f)=1/2\sqrt{f}$ at $%
D\rightarrow \infty .$ Such a weak signal can be observed if the width of
the distribution (\ref{eq42}), $D/(1+D),$ is larger than the width, $\sqrt{%
2/9N_{\pi _{tot}}}$, of the pure thermal fluctuation distribution peaked
around $f=1/3$. Unfortunately, for rare gas (when the overpopulation at the
thermal stage does not arise) one can get: $D\simeq \frac{\frac 13N_{coh}}{%
N_{gas}}\simeq \frac{D({\bf 0})}{8(R\sqrt{T_fm(t_f)})^3}\ll D({\bf 0}),$ and
so if $D({\bf 0})$ is small then the use of $f-$ distributions as signatures
of the DCC is impeded. However, at large effective sizes of the system which
could be created at RHIC and LHC, such an analysis of the $f-$distribution
turns out to be effective for a search for a weak DCC signals. Eqs. (\ref
{d-s}), (\ref{one-in}), (\ref{n-charge}) and (\ref{psi-ap}) show that for
the rare gas case, the value of $\kappa ^2(t_f)=\kappa ^2\exp (-\Gamma
_1(t_f-t_0))$ characterizes the number of coherent pions $\frac 13N_{coh}$
which, when increasing the volume, is expected to be increased also. If $%
\kappa ^2(t_f)\propto R^3\cdot const$, then $D$ is not decreasing for larger
volume while $1/\sqrt{N_\pi }$ is decreasing, for example, $\propto R^{-3/2}$
at constant phase-space densities. This fact can be used to search for a DCC
at RHIC and LHC energies even if the phase-space density is not increasing,
i.e., if the tendency for the number of pions in the central rapidity area
to be approximately proportional to the interferometry volume preserves. One
can see also that the increase in $\kappa ^2(t_f)$ helps us to distinguish a
DCC signal over the thermal pion background from spectra and correlations.

\section{Conclusions}

Some possible experimental consequences of the DCC decay have been analyzed.
Our consideration is based on the most general properties of the DCC
formation in high energy nucleus-nucleus collisions such as its localization
in a hot and dense medium, its dissipation by environment and the finite
lifetime, the existences of the freeze-out stage where the (local) thermal
equilibrium is destroyed and the misaligned DCC relaxes quickly enough to
the normal vacuum. In the picture developed, roughly speaking, a ''coherence
conservation'' low takes place: the large initial classical field (DCC)
containing virtually a plenty of coherent pions dissipates partially during
the thermal stage, but the resulting overpopulated medium amplifies the
residual weak DCC signal (''pion laser'' effect) and makes the number of
coherent particles to be large again at the freeze-out stage. Such an effect
is connected with multiboson effects in the presence of classical source and
becomes stronger when the freeze-out phase-space density increases. We have
found that the DCC signals in spectra and correlations can be observed in a
narrow energy region of soft pions, the width of this region is maximal for
the ''pion laser'' regime. The signal in the distribution of the ratio of
neutral to total pions can strongly be contaminated by the thermal
background. Nevertheless, even if the tendency of constant phase-space
densities preserves for higher energies, a possibility to detect the DCC
signal over the thermal background at so large effective volumes as expected
in RHIC and LHC heavy ion experiments has been found.

\section{Acknowledgments}

We gratefully acknowledges support by the Ukrainian State Found of the
Fundamental Research under Contract No F5.1798-98 and by Ukrainian -
Hungarian Grant No 2M/125-99.

\end{document}